\documentclass[10pt]{article}
\def\prd{Phys.~Rev.~D}
\def\prl{Phys.~Rev.~Lett. }
\def\ctg{\mathop{\rm ctg}}

\begin{document}

\title{Gravitational waveforms for spinning compact binaries}
\author{J\'anos Maj\'ar and M\'aty\'as Vas\'uth
\footnote{Electronic addresses: {\tt majar@rmki.kfki.hu, vasuth@rmki.kfki.hu}} \\ 
{\em\small KFKI Research Institute for Particle and Nuclear Physics,}\\
{\em\small Budapest 114, P.O.Box 49, H-1525 Hungary}}
%\date{\today}
\maketitle

\begin{abstract}
The rotation of the bodies and the eccentricity of the orbit have significant effects on the emitted gravitational %%@
radiation of binary systems. This work focuses on the evaluation of the gravitational wave polarization states for %%@
spinning compact binaries. We consider binaries on eccentric orbits and the spin-orbit interaction up to the 1.5 %%@
post-Newtonian order in a way which is independent of the parameterization of the orbit. The equations of motion for %%@
angular variables are included. The formal expressions of the polarization states are given with the inclusion of %%@
higher order corrections to the waveform. 
\end{abstract}

\section{Introduction}
With the construction of the ground based gravitational wave observatories, i.e. LIGO \cite{LIGO}, VIRGO 
\cite{VIRGO}, GEO600 \cite{GEO600} and TAMA \cite{TAMA}, the investigation of gravitational radiation has been %%@
receiving new impetus in the last two decades. The description of the dynamics of different astrophysical sources and %%@
the detectable signal of the emitted gravitational waves has become one of the main lines of research in general %%@
relativity and astrophysics.

Among the plausible sources of gravitational waves which can be detected by these observatories are binary systems of %%@
compact objects. Having the appropriate amplitude and frequency properties for their detection, ground based %%@
detectors are sensitive mainly for the radiation emitted in the merger era. These kinds of signals are described both %%@
by numerical simulations \cite{Num} and analytic approaches \cite{EOB1,EOB2}. Beyond the Lazarus approach %%@
\cite{L1,L2} which combines numerical techniques and perturbation theory to extract approximate waveforms, recent %%@
progress in numerical relativity have made it possible to simulate the last stages of inspiral, merger and ringdown %%@
in full general relativity \cite{P1,P2}. The comparison of numerical and analytical models shows good agreement in %%@
the gravitational wave phase evolution in the two approaches \cite{HHBG,Boyle} and helps us to check the accuracy of %%@
the analytical models and to discriminate between them \cite{BCP}.

Furthermore, for the detection of gravitational waves the description of the inspiral phase of the coalescence is %%@
also relevant. The main purpose of the determination of the detectable gravitational wave signal in the inspiral era %%@
is the LISA project \cite{LISA,LISA2}, which aims to install a detector system into space. Because of their different %%@
size and technology LISA and the other space based detectors will be most sensitive in the frequency range of the %%@
emitted radiation of inspiralling compact binaries. Because of the short duration of merger it is important to %%@
determine the direction of the source as soon as possible \cite{KGM}. An additional advantage of this description is %%@
to give analytic initial conditions for the simulations modeling the coalescence. 

During inspiral the dynamics of the source and the emitted gravitational radiation are described by two highly %%@
related methods. The post-Newtonian (PN) approximation \cite{BDI,Blanchet} is used to investigate the dynamics and %%@
the motion of the source. The essence of this approach is the neglection of ultrarelativistic and extreme %%@
gravitational effects with the introduction of a series expansion, where the expansion parameter is $\epsilon\sim %%@
v^2\sim M/r$ with the typical velocity $v$, size $r$ and total mass $M$ of the system. This approximation can be used %%@
until the binary reaches the innermost stable circular orbit \cite{BCV2,Bcirc}. To describe the emitted gravitational %%@
waves the post-Minkowskian approximation is applied. In the outer region of the source and the near zone of the %%@
radiation the two expansions can be matched together \cite{BlanchetCQG}.

The dynamics of the binary is well described in the post-Newtonian approximation. The conservative part of the %%@
dynamics is known to high post-Newtonian orders \cite{BlanchetLR}. The equations of motion with the inclusion of %%@
spin-orbit effects and the spin precession equations are given in \cite{FBB} beyond the leading-order contributions, %%@
up to 2.5\,PN order. The radial motion and the parameterization of the orbit are known up to 3\,PN order \cite{Gop}. %%@
However, these results are valid under the assumptions that the rotation of one or both of the objects and radiation %%@
reaction are negligible. For spinning compact binaries the dynamics and radiation reaction effects are discussed up %%@
to 1.5\,PN order \cite{GPV3}. The full description of motion which contains the evolution of the angular variables is %%@
known only in special cases, namely, in the test particle limit \cite{GPV1}, for binaries with one spinning %%@
components \cite{MV2} up to 1.5\,PN order, or with the neglection of spin effects up to 3\,PN order \cite{MGS}.

As a result of the gravitational wave research the transverse-traceless part of the emitted radiation field is given %%@
formally in terms of the main dynamical quantities of the binary system (the separation, relative velocity and spin %%@
vectors), the mass of the bodies and the direction of the line of sight. The results including spin effects are valid %%@
up to 1.5--2\,PN order \cite{Kidder,WW}. Higher order contributions to the waveform are known in the non-spinning and %%@
circular orbit case up to 3--3.5\,PN order \cite{BlanchetLR}. A detailed investigation of the relativistic %%@
post-Newtonian corrections with the inclusion of higher multipole moments for eccentric orbits is given in \cite{GI}. 

In our previous works we have given the description of the dynamics and a method to evaluate the detectable waveform %%@
of the emitted gravitational radiation for binaries with one spinning component and in the test particle limit %%@
\cite{MV2,MV}. Since the detectable gravitational wave signal can be unambiguously decomposed into the independent %%@
polarization states $h_+$ and $h_{\times}$ as
\begin{eqnarray}\label{h}
h(t)=F_+h_+(t)+F_{\times}h_{\times}(t)\ ,
\end{eqnarray}
where $F_+$ and $F_{\times}$ are the beam pattern functions \cite{ACST}, the main goal of our work is to construct a %%@
consistent method to determine the explicit form of these polarization states.

In the present paper we generalize our former results \cite{MV2} to discuss binary systems with components of %%@
arbitrary masses $m_1$ and $m_2$ and spins ${\bf S}_1$ and ${\bf S}_2$. Our method and formulae are valid up to %%@
1.5\,PN order and applicable to eccentric orbits independently from the parameterization of the orbit.

This paper is organized as follows. In Sec. II we present the description of the elements of the dynamics used in the %%@
projection method. In Subsection II A we introduce the main coordinate systems and the angular variables. Subsection %%@
II B contains the solution of the spin precession equations. In Subsection II C we discuss the equations for the %%@
components of the relative velocity vector and the equations of motion for the angular variables. Sec. III contains %%@
the main results of our paper, namely, the projection method and the expressions of the polarization states of the %%@
detectable gravitational waves. First we give the components of the vector triad used in the projection in Subsection %%@
III A. In Subsection III B we give the formal expressions of the polarization states in terms of the main vector %%@
components and constants of motion. Here we outline the main steps of the method which leads to the polarization %%@
states. In Sec. IV we discuss the circular orbit limit, where the explicit time dependence of the dynamical variables %%@
and the waveform can be given. In the last section we summarize our results and give a detailed outlook.

\section{The motion of the binary system}

To describe the dynamics of a binary system and the polarization states of the emitted gravitational waves one needs %%@
two different coordinate systems. The equations for the orbital elements can be easily evaluated in the invariant %%@
system which does not change in time, moreover, the projection method leads to simple expressions in the comoving %%@
system.

The following description of the evolution of the binary provides us the equations of motion for the angular %%@
variables describing the direction of the separation vector ${\bf r}$ and gives short expressions for the different %%@
components of the relative velocity vector ${\bf v}$. All these quantities depend on two scalar and one vector %%@
constants of the motion, i.e. the energy $E$ of the system, the length of the angular momentum vector ${\bf L}={\bf %%@
L}_N+{\bf L}_{PN}+{\bf L}_{SO}$, where ${\bf L}_N=\mu{\bf r}\times{\bf v}$ is the Newtonian angular momentum, %%@
$\mu=m_1m_2/M$ is the reduced mass, $M=m_1+m_2$ is the total mass of the binary, and ${\bf L}_{PN}$ and ${\bf %%@
L}_{SO}$ are the 1\,PN and spin-orbit contributions, respectively. The constant vector of motion is the total angular %%@
momentum ${\bf J}={\bf L}+{\bf S}_1+{\bf S}_2$. All these quantities and the length of the spin vectors are constant %%@
up to 2\,PN order.

\subsection{Angular variables}

The dynamics of the binary and the evolution of the orbital elements are most conveniently described in the invariant %%@
coordinate system. The $z$ axis of this system is fixed to the direction of the total angular momentum vector ${\bf %%@
J}$. There is another constant vector in our description, the direction of the line of sight ${\bf N}$. To simplify %%@
the calculations the $x$ and $y$ axes of the invariant system are chosen in a way that ${\bf %%@
N}=\left(\sin{\gamma},0,\cos{\gamma}\right)$ in this system with the constant angle $\gamma$ of ${\bf J}$ and ${\bf %%@
N}$.

The projection method results in simple expressions for the polarization states if we express the components of the %%@
main vector quantities and the transverse traceless part $h^{ij}_{TT}$ of the radiation field in the comoving %%@
coordinate system. The $x$ and $z$ axes of this coordinate system are aligned with the separation vector ${\bf r}$ %%@
and the direction of the Newtonian angular momentum ${\bf L}_N$, respectively.

The transformation between these coordinate systems is characterized by Euler-angles \cite{GPV2}. The vector ${\bf %%@
u}$ in the comoving system is parameterized as
\begin{eqnarray}
{\bf u}'=R_z(\Phi)R_x(\iota)R_z(\Psi){\bf u} \label{trans}
\end{eqnarray}
in the invariant system, where $\iota$ is the angle between ${\bf J}$ and ${\bf L}_N$, $\Phi$ describes the %%@
precession of the orbit and $\Psi$ is the polar angle on the orbital plane. Using this transformation the separation %%@
vector in the invariant system becomes
\begin{eqnarray}
{\bf r}=r\left(\begin{array}{c} \cos{\Phi}\cos{\Psi}-\cos{\iota}\sin{\Phi}\sin{\Psi} \\
\sin{\Phi}\cos{\Psi}+\cos{\iota}\cos{\Phi}\sin{\Psi} \\
\sin{\iota}\sin{\Psi}\end{array} \right)\ .
\end{eqnarray}
In the comoving system the components of the relative velocity vector are expressed by Euler-angles,
\begin{eqnarray}
{\bf v}=\left(\begin{array}{c} \dot{r} \\
r\left(\cos{\iota}\dot{\Phi}+\dot{\Psi}\right) \\
0 \end{array}\right)\label{vvecdiff}\ .
\end{eqnarray}
For further convenience we decompose the relative velocity vector as
\begin{eqnarray}
{\bf v}=v_{\parallel}{\bf n}+v_{\perp}{\bf m}\label{vvecdec}\ ,
\end{eqnarray}
where ${\bf n}={\bf r}/r$ is the direction of the separation vector and ${\bf m}$ is the unit vector in the direction %%@
of the $y$ axis of the comoving system. Moreover, to separate terms which contribute to different PN orders and the %%@
rotation of the bodies we decompose the main quantities as follows
\begin{eqnarray}\label{decomp}
r=r_N+r_{PN}+r_S\ ,\qquad
v_{\parallel}=v_{\parallel N}+v_{\parallel PN}+v_{\parallel S}\ ,\qquad
v_{\perp}=v_{\perp N}+v_{\perp PN}+v_{\perp S}\ ,\nonumber\\
\Psi=\Psi_N+\Psi_{PN}+\Psi_S\ ,\qquad
\Phi=\Phi_N+\Phi_{PN}+\Phi_S\ ,\qquad
\iota=\iota_N+\iota_{PN}+\iota_S\ ,
\end{eqnarray}
where $r$ denotes the length of the separation vector.

During the calculation of the polarization states we do not need the expressions for all the Euler angles separately. %%@
As it will be shown below only the combinations $\Upsilon=\Phi+\Psi$, $\iota_S\sin{\Phi_N}$ and $\iota_S\cos{\Phi_N}$ %%@
appear. To determine the equations for these quantities we note that $\iota_N=0$, see the Appendix of \cite{GPV2}. %%@
Since the 1\,PN corrections do not change the direction of the orbital plane $\iota_{PN}=0$, and hence %%@
$\iota=\iota_S$.

\subsection{Spin-precession}

The first step in describing the dynamics of the binary system is the examination of the spin-precession equations,
\begin{eqnarray}
\dot{\bf S}_1&=&\frac{1}{r^3}\left(\frac{4+3\zeta_1}{2}\,{\bf L}_N-{\bf S}_2
+\frac{3}{r^2}({\bf r{\mbox{\boldmath$\cdot$}}S}_2)\,{\bf r}\right)\times{\bf S}_1\ , \\
\dot{\bf S}_2&=&\frac{1}{r^3}\left(\frac{4+3\zeta_2}{2}\,{\bf L}_N-{\bf S}_1
+\frac{3}{r^2}({\bf r{\mbox{\boldmath$\cdot$}}S}_1)\,{\bf r}\right)\times{\bf S}_2\ ,
\end{eqnarray}
where $\zeta_1=\zeta_2^{-1}={m_2}/{m_1}$.

Since we are interested in the lowest order contributions (these contributions give rise to 1.5\,PN effects in the %%@
description of motion) the spin-spin interaction which appears at higher order is neglected and hence
\begin{eqnarray}
\dot{\bf S}_i=\frac{(4+3\zeta_i)}{2r^3}\,{\bf J}\times{\bf S}_i
\end{eqnarray}
for $i=1,2$.

The components of the spin vectors are expressed with the magnitude $S_i$ and the spherical angles ${\alpha_i}$ and %%@
${\beta_i}$. The evolution of these angles is governed by
\begin{eqnarray}
S_i\left(\begin{array}{c}
-\sin{\beta_i}\sin{\alpha_i}\dot{\beta}_i+\cos{\beta_i}\cos{\alpha_i}\dot{\alpha}_i \\
\cos{\beta_i}\sin{\alpha_i}\dot{\beta}_i+\sin{\beta_i}\cos{\alpha_i}\dot{\alpha}_i \\
\sin{\alpha_i}\dot{\alpha}_i
\end{array} \right)
=(4+3\zeta_i)\frac{JS_i}{2r^3}\left(\begin{array}{c}
-\sin{\beta_i}\sin{\alpha_i} \\
\cos{\beta_i}\sin{\alpha_i} \\
0
\end{array} \right)\ ,
\end{eqnarray}
where the lengths of the spin vectors change only under radiation effects over 2\,PN order. In this way we get the %%@
following equations for the spherical angles
\begin{eqnarray}
\dot{\alpha}_i=0\ , \quad
\dot{\beta}_i=(4+3\zeta_i)\frac{J}{2r^3}\label{spinprang}\ .
\end{eqnarray}

In the Newtonian case $\Upsilon_N$ governs the angular dynamics of the system. With the use of Eq. (\ref{spinprang}) %%@
we relate the relative PN order of the angular variables to it, 
\begin{eqnarray}
\frac{\dot{\beta}_i}{\dot{\Upsilon}_N}=\frac{(4+3\zeta_i)J/(2r_N^3)}{L/(\mu r_N^2)} \sim\frac{L/(r_N^3)}{L/(\mu %%@
r_N^2)}=\frac{\mu}{r_N}\sim\epsilon\ .
\end{eqnarray}
Since we require that the integration of the differential equations does not change the relative PN order of the %%@
different quantities the same estimate is right for the angular variables, henceforth $\beta_i / %%@
\Upsilon_N\sim\epsilon$. This statement indicates that we can introduce the decomposition %%@
$\beta_i=\beta_{iN}+\beta_{iPN}$, where the relative order of these angles are 
${\beta_{iN}}/{\Upsilon_N}\sim 1$ and ${\beta_{iPN}}/{\Upsilon_N}\sim\epsilon$.
Here $\beta_{iN}$ are constant and
\begin{eqnarray}
\dot{\beta}_{iPN}&=&(4+3\zeta_i)\frac{J}{2r_N^3}\label{betaipn}\ .
\end{eqnarray}

\subsection{The equations of motion}

If one combines the results of \cite{GPV3} and \cite{KMG} the length and the components of the relative velocity %%@
vector can be written as
\begin{eqnarray}\label{veloc}
v^2&=&\frac{2E}{\mu}+\frac{2M}{r}-\left[3(1-3\eta)\left(\frac{E}{\mu}\right)^2
+2(6-7\eta)\frac{EM}{\mu r}+(10-5\eta)\left(\frac{M}{r}\right)^2-\frac{\eta ML^2}{\mu^2r^3}\right]
-\frac{2({\bf L}{\mbox{\boldmath$\cdot\sigma$}})}{\mu r^3}\ ,\nonumber\\
v_{\parallel}^2&=&\dot{r}^2=\frac{2E}{\mu}+\frac{2M}{r}-\frac{L^2}{\mu^2r^2}-\left[3(1-3\eta)\left(\frac{E}{\mu}
\right)^2+ 2(6-7\eta)\frac{EM}{\mu r}
+(10-5\eta)\left(\frac{M}{r}\right)^2-(2-6\eta)\frac{EL^2}{\mu^3r^2}\right.\nonumber\\
&-&\left.(8-3\eta)\frac{ML^2}{\mu^2r^3}\right]+\frac{2E({\bf L{\mbox{\boldmath$\cdot\sigma$}}})}{M\mu^2r^2}
-\frac{2}{\mu r^3}(2{\bf L{\mbox{\boldmath$\cdot$}}S}+{\bf L{\mbox{\boldmath$\cdot\sigma$}}})\ ,
\end{eqnarray}
where $\eta=\mu/M$ and we use the notations 
${\mbox{\boldmath$\sigma$}}=\zeta_1{\bf S}_1+\zeta_2{\bf S}_2$ and ${\bf S}={\bf S}_1+{\bf S}_2$. 
Since $v_{\perp}^2=v^2-\dot{r}^2$ we have
\begin{eqnarray}\label{vperp}
v_{\perp}=\frac{L}{\mu r_N}
-\frac{L}{\mu r_N}\left[\frac{r_{PN}}{r_N}+(1-3\eta)\frac{E}{\mu}+(4-2\eta)\frac{M}{r_N}\right]
-\frac{L}{\mu r_N}\left[\frac{r_S}{r_N}+\frac{E({\bf L{\mbox{\boldmath$\cdot\sigma$}}})}{L^2M}
-\frac{2\mu({\bf L{\mbox{\boldmath$\cdot$}}S})}{L^2r_N}\right]\ .
\end{eqnarray}

To derive the equation for the angle $\Upsilon$ we use the form of the relative velocity vector given in Eq.  %%@
(\ref{vvecdiff}) and that $\iota=\iota_S$. From the second component we have $v_{\perp}=r\dot{\Upsilon}$, and hence
\begin{eqnarray}\label{upsilon}
\dot{\Upsilon}_N&=&\frac{L}{\mu r_N^2}\ ,\qquad
\dot{\Upsilon}_{PN}=-\frac{L}{\mu r_N^2}\left[\frac{2r_{PN}}{r_N}+(1-3\eta)\frac{E}{\mu}
+(4-2\eta)\frac{M}{r_N}\right]\ ,\nonumber\\
\dot{\Upsilon}_S&=&-\frac{L}{\mu r_N^2}\left[\frac{2r_S}{r_N}
+\frac{E({\bf L{\mbox{\boldmath$\cdot\sigma$}}})}{L^2M}
-\frac{2\mu({\bf L{\mbox{\boldmath$\cdot$}}S})}{L^2r_N}\right]\ .
\end{eqnarray}

To evaluate the remaining equations the components of the total angular momentum ${\bf J}={\bf L}_{N}+{\bf %%@
L}_{PN}+{\bf L}_{SO}+{\bf S}$ are determined, where
\begin{eqnarray}
{\bf L}_{PN}={\bf L}_N\left[\frac{1-3\eta}{2}\,v^2+(3+\eta)\frac{M}{r}\right]
\end{eqnarray}
is the 1\,PN correction of the angular momentum vector and
\begin{eqnarray}
{\bf L}_{SO}=\eta\left\{\frac{M}{r^3}[{\bf r}\times({\bf r}\times(2{\bf S}+{\mbox{\boldmath$\sigma$}}))]
-\frac{1}{2}[{\bf v}\times({\bf v}\times{\mbox{\boldmath$\sigma$}})]\right\}
\end{eqnarray}
is the spin-orbit contribution. In this way the components of ${\bf J}$ are
\begin{eqnarray}
{\bf J}=\left(\begin{array}{c} 
L\iota_S\sin{\Phi_N}+\displaystyle\sum_{i=1}^2\left[A_i\sin{\xi_i}+B_i\cos{\xi_i}+(C_i+1)\cos{\beta_{iN}}
+\beta_{iPN}\sin{\beta_{iN}}\right] S_i\sin{\alpha_i}\\
-L\iota_S\cos{\Phi_N}+\displaystyle\sum_{i=1}^2\left[B_i\sin{\xi_i}-A_i\cos{\xi_i}+(C_i+1)\sin{\beta_{iN}}
-\beta_{iPN}\cos{\beta_{iN}}\right] S_i\sin{\alpha_i}\\
L+\displaystyle\sum_{i=1}^2(C_i+1)S_i\cos{\alpha_i}
\end{array} \right)\ ,
\end{eqnarray}
where we have introduced the notation $\xi_i=2\Upsilon_N-\beta_i$ and
\begin{eqnarray}
A_i=\frac{\eta\zeta_iv_{\parallel N}v_{\perp N}}{2}\ , \quad
B_i=\left(\frac{(2+\zeta_i)\mu}{2r_N}-\frac{\eta\zeta_iv_{\parallel N}^2}{4}
+\frac{\eta\zeta_iv_{\perp N}^2}{4}\right)\ , \quad
C_i=\left(-\frac{(2+\zeta_i)\mu}{2r_N}+\frac{\eta\zeta_iv_N^2}{4}\right)\ .
\end{eqnarray}

Since the total angular momentum vector is ${\bf J}=(0,0,J)$ in the invariant system these equations give rise to the %%@
identities
\begin{eqnarray}
\sum_{i=1}^2S_i\cos{\beta_{iN}}\sin{\alpha_i}=0\ ,\quad
\sum_{i=1}^2S_i\sin{\beta_{iN}}\sin{\alpha_i}=0
\end{eqnarray}
in 0.5\,PN order, and 
\begin{eqnarray}\label{iophi}
\iota_S\sin{\Phi_N}&=&-\frac1L\sum_{i=1}^2\left(A_i\sin{\xi_i}+B_i\cos{\xi_i}+C_i\cos{\beta_{iN}}+ %%@
\beta_{iPN}\sin{\beta_{iN}}\right)S_i\sin{\alpha_i}\ ,\\ 
\iota_S\cos{\Phi_N}&=&\frac1L\sum_{i=1}^2\left(B_i\sin{\xi_i}-A_i\cos{\xi_i}+C_i\sin{\beta_{iN}}- %%@
\beta_{iPN}\cos{\beta_{iN}}\right)S_i\sin{\alpha_i}\nonumber
\end{eqnarray}
in 1.5\,PN order. As it is not necessary to evaluate $\iota_S$ and $\Phi_N$ separately with Eqs. (\ref{iophi}) we %%@
have all the expressions required for the description of motion of the binary system and the emitted gravitational %%@
radiation.

\section{The polarization states}

\subsection{The projection}

The polarization states of the detectable gravitational waves are determined by the projections of the %%@
transverse-traceless tensor $h^{ij}_{TT}$ representing metric perturbations,
\begin{eqnarray}
h_+=\frac{1}{2}(p_ip_j-q_iq_j)h^{ij}_{TT}\ ,\quad
h_{\times}=\frac{1}{2}(p_iq_j+q_ip_j)h^{ij}_{TT} \ ,\label{h-k}
\end{eqnarray}
where ${\bf p}$ is a unit vector in the orbital plane perpendicular to the direction of the line of sight ${\bf N}$, %%@
and ${\bf q}={\bf N}\times{\bf p}$.

Since we have chosen the comoving system to describe this projection we determine the components of ${\bf N}$, ${\bf %%@
p}$ and ${\bf q}$ in this system. With the use of the transformation law (\ref{trans}) ${\bf N}$ has the following %%@
form in the comoving system  
\begin{eqnarray}\label{nvec}
{\bf N}=\left(\begin{array}{c}
\sin{\gamma}\cos{\Upsilon_N}-\left[(\Upsilon_{PN}+\Upsilon_S)\sin{\gamma}\sin{\Upsilon_N}- %%@
\cos{\gamma}\sin{\Phi_N}\iota_S\right]\\
-\sin{\gamma}\sin{\Upsilon_N}-\left[(\Upsilon_{PN}+\Upsilon_S)\sin{\gamma}\cos{\Upsilon_N}- %%@
\cos{\gamma}\cos{\Phi_N}\iota_S\right]\\
\cos{\gamma}+\sin{\gamma}\sin{\Phi_N}\iota_S
\end{array}\right)\ .
\end{eqnarray}
Moreover, the conditions for ${\bf p}$ determine its components as
\begin{eqnarray}\label{pvec}
{\bf p}=\left(\begin{array}{c}
\sin{\Upsilon_N}+(\Upsilon_{PN}+\Upsilon_S-\ctg{\gamma}\cos{\Phi_N}\iota_S)\cos\Upsilon_N\\
\cos{\Upsilon_N}-(\Upsilon_{PN}+\Upsilon_S-\ctg{\gamma}\cos{\Phi_N}\iota_S)\sin\Upsilon_N\\
0
\end{array}\right)\ .
\end{eqnarray}
From Eqs. (\ref{nvec}) and (\ref{pvec}) it follows that for the description of the polarization states the relevant %%@
angular variables are $\Upsilon$ and the combinations $\iota_S\cos{\Phi_N}$ and $\iota_S\sin{\Phi_N}$.

\subsection{The expressions for the polarization states}

To separate the different contributions the transverse-traceless part of the radiation field $h^{ij}_{TT}$ can be %%@
decomposed in the post-Newtonian approximation as \cite{Kidder}
\begin{eqnarray}
h^{ij}_{TT}=\frac{2\mu}{D}\left[Q^{ij}+P^{0.5}Q^{ij}+PQ^{ij}+PQ^{ij}_{SO}+P^{1.5}Q^{ij} 
+P^{1.5}Q^{ij}_{SO}\right]_{TT}\ ,\label{hTT}
\end{eqnarray}
where $D$ is the distance between the observer and the source and we have collected all the terms which are relevant %%@
up to 1.5\,PN order. $Q^{ij}$ denotes the quadrupole (or Newtonian) term, $P^{0.5}Q^{ij}$, $PQ^{ij}$ and %%@
$P^{1.5}Q^{ij}$ are higher order relativistic corrections and $PQ^{ij}_{SO}$ and $P^{1.5}Q^{ij}_{SO}$ are the %%@
spin-orbit terms. The detailed expressions for these contributions are given in \cite{Kidder,WW}.

We choose a similar decomposition for the polarization states $h_+$ and $h_{\times}$,
\begin{eqnarray}\label{hdecomp}
h_{^+_{\times}}=\frac{2\mu}{D}\left[h_{^+_{\times}}{}^N+h_{^+_{\times}}{}^{0.5}
+ h_{^+_{\times}}{}^1+h_{^+_{\times}}{}^{1SO}+h_{^+_{\times}}{}^{1.5}+ h_{^+_{\times}}{}^{1.5SO}\right]\ .
\end{eqnarray}
It is worth to mention that this decomposition is formal in the sense that higher order terms in the description of %%@
motion can cause higher order contributions in $h_{^+_{\times}}{}^N$ and $h_{^+_{\times}}{}^{0.5}$.

With the use of the decomposition (\ref{vvecdec}) of the relative velocity vector we evaluate the projections in Eq. %%@
(\ref{h-k}) and obtain the following formulae for the "plus" polarization state
\begin{eqnarray}\label{hplus}
h_+^N&=&\left(\dot{r}^2-\frac{M}{r}\right)(p_x^2-q_x^2)+2v_{\perp}\dot{r}(p_xp_y-q_xq_y)
+ v_{\perp}^2(p_y^2-q_y^2)\ ,\nonumber\\
h_+^{0,5}&=&\frac{\delta m}{M}\left(\left(\dot{r}\left[\frac{2M}{r}-\frac{\dot{r}^2}{2}\right]N_x
+ v_{\perp}\left[\frac{M}{2r}-\dot{r}^2\right]N_y\right)(p_x^2-q_x^2)\right.\nonumber\\ 
&+&v_{\perp}\left(\left[\frac{3M}{r}-2\dot{r}^2\right]N_x-2v_{\perp}\dot{r}
N_y\right)(p_xp_y-q_xq_y)\nonumber\\
&-&\left.v_{\perp}^2\left(\dot{r}N_x+v_{\perp}N_y\right)(p_y^2-q_y^2)\right)\ ,\nonumber\\
h_+^1&=&\frac{1}{6}\left((1-3\eta)\left[\left(-\frac{21\dot{r}^2M}{r}+\frac{3Mv^2}{r}+6\dot{r}^4 
+\frac{7M^2}{r^2}\right)N_x^2+ 4v_{\perp}\dot{r}\left(-\frac{6M}{r}+3\dot{r}^2\right)N_xN_y\right.\right.\nonumber\\
&+&\left.\left.2v_{\perp}^2\left(3\dot{r}^2-\frac{M}{r}\right)N_y^2\right]
+ \left[\frac{(19-9\eta)\dot{r}^2M}{r}+(3-9\eta)v^2\dot{r}^2-\frac{(10+3\eta)v^2M}{r}+\frac{29M^2}{r^2}\right]\right) 
(p_x^2-q_x^2)\nonumber\\
&+&\frac{v_{\perp}}{6}\left((1-3\eta)\left[6\dot{r}\left(-\frac{5M}{r}+2\dot{r}^2\right)N_x^2 
+8v_{\perp}\left(-4\frac{M}{r}+3\dot{r}^2\right)N_xN_y+ 12v_{\perp}^2\dot{r}N_y^2\right]\right.\nonumber\\
&+&\left.6\dot{r}\left[\frac{(2+4\eta)M}{r}+(1-3\eta)v^2\right]\right)(p_xp_y-q_xq_y)\nonumber\\ 
&+&\frac{v_{\perp}^2}{6}\left((1-3\eta)\left[2\left(-\frac{7M}{r}+3\dot{r}^2\right)N_x^2
+12v_{\perp}\dot{r}N_xN_y+6v_{\perp}^2N_y^2\right]\right.\nonumber\\
&+&\left.\left[-\frac{(4-6\eta)M}{r}+ (3-9\eta)v^2\right]\right)(p_y^2-q_y^2)\ ,\nonumber\\
h_+^{1SO}&=&-\frac{1}{r^2}\left[({\bf \Delta{\mbox{\boldmath$\cdot$}}q})p_x
+({\bf \Delta{\mbox{\boldmath$\cdot$}}p})q_x\right]\ ,\nonumber\\
h_+^{1,5}&=&\frac{\delta m}{M}\left\{(1-2\eta)\left(\dot{r}\left[\frac{3\dot{r}^2M}{4r}
-\frac{v^2M}{r}- \frac{41M^2}{12r^2}-\dot{r}^4\right]N_x^3\right.\right.\nonumber\\
&+&\left.v_{\perp}\left[\frac{85\dot{r}^2M}{8r}-\frac{9v^2M}{8r}
- \frac{7M^2}{2r^2}-3\dot{r}^4\right]N_x^2N_y+3\dot{r}v_{\perp}^2\left[\frac{2M}{r}- \dot{r}^2\right]N_xN_y^2
+v_{\perp}^3\left[\frac{M}{4r}-\dot{r}^2\right]N_y^3\right)\nonumber\\
&+&\dot{r}\left[-\frac{(10+7\eta)\dot{r}^2M}{2r}+\frac{(2+\eta)v^2M}{2r}
-\frac{(59-30\eta)M^2}{12r^2}-\frac{(1-5\eta)v^2\dot{r}^2}{2}\right]N_x\nonumber\\
&+&\left.v_{\perp}\left[-\frac{(25+26\eta)\dot{r}^2M}{8r}+\frac{(7-2\eta)v^2M}{8r}- \frac{(26-3\eta)M^2}{6r^2}
-\frac{(1-5\eta)v^2\dot{r}^2}{2}\right]N_y\right\}(p_x^2-q_x^2)\nonumber\\
&+&v_{\perp}\frac{\delta m}{M}\left\{(1-2\eta)\left(\left[\frac{\dot{r}^2M}{4r}-\frac{7v^2M}{4r}
-\frac{11M^2}{r^2}-2\dot{r}^4\right]N_x^3+v_{\perp}\dot{r}\left[\frac{16M}{r}
-6\dot{r}^2\right]N_x^2N_y\right.\right.\nonumber\\
&+&\left.3v_{\perp}^2\left[\frac{5M}{2r}-2\dot{r}^2\right]N_xN_y^2-2v_{\perp}^3\dot{r}N_y^3\right) \nonumber\\
&+&\left[-\frac{(49+14\eta)\dot{r}^2M}{4r}+\frac{(11-6\eta)v^2M}{4r}-\frac{(32-9\eta)M^2}{3r^2}
-(1-5\eta)v^2\dot{r}^2\right]N_x\nonumber\\
&-&\left.v_{\perp}\dot{r}\left[\frac{(2+6\eta)M}{r}+ (1-5\eta)v^2\right]N_y\right\}(p_xp_y-q_xq_y)\nonumber\\
&+&v_{\perp}^2\frac{\delta m}{M}\left\{(1-2\eta)\left(-\dot{r}\left[\frac{5M}{4r}+\dot{r}^2\right]N_x^3
+v_{\perp}\left[\frac{29M}{4r}-3\dot{r}^2\right]N_x^2N_y-3v_{\perp}^2\dot{r}N_xN_y^2
-v_{\perp}^3N_y^3\right)\right.\nonumber\\
&-&\left.\dot{r}\left[\frac{(7+3\eta)M}{r}+\frac{(1-5\eta)v^2}{2}\right]N_x+ v_{\perp}\left[\frac{(3-8\eta)M}{4r}
-\frac{(1-5\eta)v^2}{2}\right]N_y\right\}(p_y^2-q_y^2)\ ,\nonumber\\
h_+^{1,5SO}&=&\frac{2}{r^2}\left\{3v_{\perp}(2S_z+\Delta_z)(p_x^2-q_x^2)
+\frac{\dot{r}}{2}[({\bf S}-{\bf\Delta})\times(p_x{\bf p}-q_x{\bf q})]_x
-\frac{v_{\perp}}{2}[(9{\bf S}+5{\bf\Delta})\times(p_x{\bf p}-q_x{\bf q})]_y\right.\nonumber\\
&-&\left.v_{\perp}[({\bf S}+{\bf\Delta})\!\times\!(p_y{\bf p}-q_y{\bf q})]_x
\!-\!({\bf S}+{\bf\Delta}){\mbox{\boldmath$\cdot$}}\left[\left(\frac{\dot{r}}{2}N_x\!+\!v_{\perp}N_y\right)
(p_x{\bf q}+q_x{\bf p})\!+\!v_{\perp}N_x(p_y{\bf q}+q_y{\bf p})\right]\right\},
\end{eqnarray}
and similarly for the "cross" polarization state
\begin{eqnarray}\label{hcross}
h_{\times}^N&=&2\left(\left(\dot{r}^2-\frac{M}{r}\right)p_xq_x+v_{\perp}\dot{r}(p_xq_y+q_xp_y) 
+v_{\perp}^2p_yq_y\right)\ ,\nonumber\\
h_{\times}^{0,5}&=&\frac{\delta m}{M}\left(\dot{r}\left[\frac{4M}{r}-\dot{r}^2\right]N_x+v_{\perp}\left[\frac{M}{r}
-2\dot{r}^2\right]N_y\right)p_xq_x\nonumber\\ &+&v_{\perp}\left(\left[\frac{3M}{2r}-\dot{r}^2\right]N_x
-v_{\perp}\dot{r}N_y\right)(p_xq_y+q_xp_y)\nonumber\\
&-&2v_{\perp}^2\left(\dot{r}N_x+v_{\perp}N_y\right)p_yq_y\ ,\nonumber\\
h_{\times}^{1}&=&\frac{1}{3}\left((1-3\eta)\left(\left[-\frac{21\dot{r}^2M}{r}+\frac{3Mv^2}{r}+ 6\dot{r}^4 
+\frac{7M^2}{r^2}\right]N_x^2+ 4v_{\perp}\dot{r}\left[-\frac{6M}{r}+3\dot{r}^2\right]N_xN_y\right.\right.\nonumber\\
&+&\left.\left.2v_{\perp}^2\left[3\dot{r}^2-\frac{M}{r}\right]N_y^2\right)
+\left[\frac{(19-9\eta)\dot{r}^2M}{r}+(3-9\eta)v^2\dot{r}^2-\frac{(10+3\eta)v^2M}{r}
+\frac{29M^2}{r^2}\right] \right) p_xq_x\nonumber\\
&+&\frac{v_{\perp}}{6}\left((1-3\eta)\left(6\dot{r}\left[-\frac{5M}{r}+2\dot{r}^2\right]N_x^2 
+8v_{\perp}\left[-4\frac{M}{r}+3\dot{r}^2\right]N_xN_y+ 12v_{\perp}^2\dot{r}N_y^2\right)\right.\nonumber\\
&+&\left.6\dot{r}\left[\frac{(2+4\eta)M}{r}+(1-3\eta)v^2\right]\right)(p_xq_y+q_xp_y)\nonumber\\ 
&+&\frac{v_{\perp}^2}{3}\left((1-3\eta)\left(2\left[-\frac{7M}{r}+3\dot{r}^2\right]N_x^2
+12v_{\perp}\dot{r}N_xN_y+6v_{\perp}^2N_y^2\right)\!-\!\left[\frac{(4-6\eta)M}{r}
-(3-9\eta)v^2\right]\right)p_yq_y\ ,\nonumber\\
h_{\times}^{1SO}&=&-\frac{1}{r^2}\left[({\bf\Delta{\mbox{\boldmath$\cdot$}}q})q_x
-({\bf\Delta{\mbox{\boldmath$\cdot$}}p})p_x\right]\ ,\nonumber\\
h_{\times}^{1,5}&=&\frac{\delta m}{M}\left\{(1-2\eta)\left(\dot{r}\left[\frac{3\dot{r}^2M}{2r}-\frac{2v^2M}{r}
- \frac{41M^2}{6r^2}-2\dot{r}^4\right]N_x^3\right.\right.\nonumber\\
&+&\left.v_{\perp}\left[\frac{85\dot{r}^2M}{4r}-\frac{9v^2M}{4r}- \frac{7M^2}{r^2}-6\dot{r}^4\right]N_x^2N_y
+ 6\dot{r}v_{\perp}^2\left[\frac{2M}{r}-\dot{r}^2\right]N_xN_y^2
+v_{\perp}^3\left[\frac{M}{2r}-2\dot{r}^2\right]N_y^3\right)\nonumber\\
&+&\dot{r}\left[-\frac{(10+7\eta)\dot{r}^2M}{r}+\frac{(2+\eta)v^2M}{r}- \frac{(59-30\eta)M^2}{6r^2}
- (1-5\eta)v^2\dot{r}^2\right]N_x\nonumber\\
&+&\left.\left[-\frac{(25+26\eta)v_{\perp}\dot{r}^2M}{4r}+\frac{(7-2\eta)v^2v_{\perp}M}{4r}
- \frac{(26-3\eta)v_{\perp}M^2}{3r^2}- (1-5\eta)v^2v_{\perp}\dot{r}^2\right]N_y\right\}p_xq_x\nonumber\\
&+&v_{\perp}\frac{\delta m}{M}\left\{(1-2\eta)\left(\left[\frac{\dot{r}^2M}{4r}-\frac{7v^2M}{4r}
- \frac{11M^2}{r^2}-2\dot{r}^4\right]N_x^3+v_{\perp}\dot{r}\left[\frac{16M}{r}
- 6\dot{r}^2\right]N_x^2N_y\right.\right.\nonumber\\
&+&\left.3v_{\perp}^2\left[\frac{5M}{2r}-2\dot{r}^2\right]N_xN_y^2-2v_{\perp}^3\dot{r}N_y^3\right)
+\left[-\frac{(49+14\eta)\dot{r}^2M}{4r}+\frac{(11-6\eta)v^2M}{4r}\right.\nonumber\\
&-&\left.\left.\frac{(32-9\eta)M^2}{3r^2}-(1-5\eta)v^2\dot{r}^2\right]N_x
- v_{\perp}\dot{r}\left[\frac{(2+6\eta)M}{r}+ (1-5\eta)v^2\right]N_y\right\}(p_xq_y+q_xp_y)\nonumber\\
&+&v_{\perp}^2\frac{\delta m}{M}\left\{(1-2\eta)\left(-\dot{r}\left[\frac{5M}{2r}+\dot{r}^2\right]N_x^3
+v_{\perp}\left[\frac{29M}{2r}-6\dot{r}^2\right]N_x^2N_y-6v_{\perp}^2\dot{r}N_xN_y^2
-2v_{\perp}^3N_y^3\right)\right.\nonumber\\
&-&\left.\dot{r}\left[\frac{(14+6\eta)M}{r}+(1-5\eta)v^2\right]N_x
+ v_{\perp}\left[\frac{(3-8\eta)M}{2r}-(1-5\eta)v^2\right]N_y\right\}p_yq_y\ ,\nonumber\\
h_{\times}^{1,5SO}&=&\frac{2}{r^2}\left\{6v_{\perp}(2S_z+\Delta_z)p_xq_x
+\frac{\dot{r}}{2}[({\bf S}-{\bf\Delta})\times(p_x{\bf q}+q_x{\bf p})]_x
-\frac{v_{\perp}}{2}[(9{\bf S}+5{\bf\Delta})\times(p_x{\bf q}+q_x{\bf p})]_y\right.\nonumber\\
&-&\left.v_{\perp}[({\bf S}+{\bf\Delta})\times(p_y{\bf q}+q_y{\bf p})]_x
-({\bf S}+{\bf\Delta})\,{\mbox{\boldmath$\cdot$}}\left[\left(\frac{\dot{r}}{2}N_x+v_{\perp}N_y\right)(q_x{\bf q}
-p_x{\bf p})+v_{\perp}N_x(q_y{\bf q}-p_y{\bf p})\right]\right\}\ ,
\end{eqnarray}
where $\delta m=m_2-m_1$ and ${\bf\Delta}=M({{\bf S}_2}/{m_2}-{{\bf S}_1}/{m_1})$.

The main steps of the method, which results in the detectable gravitational wave signal, are the following. With the %%@
use of an appropriate parameterization, i.e. a solution of the radial equation, one can integrate the spin precession %%@
equations, Eq. (\ref{betaipn}). With the solution of Eqs. (\ref{veloc})-(\ref{upsilon}) and (\ref{iophi}) the %%@
parameter dependence of $v_{\parallel}$, $v_{\perp}$, $\Upsilon$ and the $\iota_S\sin{\Phi_N}$ and %%@
$\iota_S\cos{\Phi_N}$ combinations can be given.

The second step is to substitute these quantities into Eqs. (\ref{nvec}) and (\ref{pvec}) to get the components of %%@
the ${\bf N}$, ${\bf p}$, ${\bf q}$ orthonormal triad. With the use of the transformation Eq. (\ref{trans}) one can %%@
evaluate the components of the spin vectors in the comoving system.

The last step is to substitute these results into Eqs. (\ref{hdecomp})-(\ref{hcross}) and collect all the terms which %%@
belong to the same PN order. To get the explicit form of the polarization states quadratic or higher order spin %%@
contributions are neglected. The detectable wave signal arises with the use of Eq. (\ref{h}) and the explicit form of %%@
the beam-pattern functions $F_+$ and $F_{\times}$ \cite{ACST}.

\section{The circular orbit case}

To investigate the effects of the orbital eccentricity on the detectable waveform we present the results of the %%@
method outlined above in the circular orbit limit. Furthermore, as the eccentricity decreases under radiation %%@
reaction in most cases the motion of the binary becomes circular at later stages of inspiral when the highest %%@
amplitude signals are emitted. 

To define circular orbits the decomposition (\ref{vvecdec}) of the relative velocity vector is reformulated as
\begin{eqnarray}
{\bf v}=\dot{r}{\bf n}+r\omega{\bf m}
\end{eqnarray}
with $\dot{r}=\dot{\omega}=0$. The connection between the above form of ${\bf v}$ and the angular variables %%@
introduced earlier is given as
\begin{eqnarray}\label{circomega}
\omega=\dot{\Phi}+\dot{\Psi}=\dot{\Upsilon}\ ,
\end{eqnarray}
see Eq. (\ref{vvecdiff}). With the use of these quantities the description of motion and the expressions of the ${\bf %%@
N}$, ${\bf p}$, ${\bf q}$, ${\bf S}_1$ and ${\bf S}_2$ vectors become simpler and they can be integrated explicitly %%@
in time $t$. To keep the expressions of the polarization states as simple as possible we use the following form of %%@
$v^2$ given in \cite{Kidder} up to 1.5\,PN order
\begin{eqnarray}
v^2=v_{\perp}^2=r^2\omega^2=\frac{M}{r}\left[1-(3-\eta)\left(\frac{M}{r}\right)
-\frac{1}{M^2}\sum_{i=1,2}(2+3\zeta^i)
({\bf n}\times{\bf m}){\mbox{\boldmath$\cdot$}}{\bf S}_i\left(\frac{M}{r}\right)^{3/2}\right]\ .
\end{eqnarray}
In this way the contributions to $h_+$ and $h_{\times}$ can be reformulated as
\begin{eqnarray}
h_+^N&=&\left(\frac{M}{r}\right)[-(p_x^2-q_x^2)+(p_y^2-q_y^2)]\ ,\nonumber\\
h_+^{0,5}&=&\left(\frac{M}{r}\right)^{3/2}\frac{\delta m}{M}\left[\frac{1}{2}N_y(p_x^2-q_x^2)
+3N_x(p_xp_y-q_xq_y)-N_y(p_y^2-q_y^2)\right]\ ,\nonumber\\
h_+^1&=&\left(\frac{M}{r}\right)^2\frac{1}{6}\left(\left[(1-3\eta)(10N_x^2-2N_y^2)
+ (19-3\eta)\right](p_x^2-q_x^2)-32(1-3\eta)N_xN_y(p_xp_y-q_xq_y)\right.\nonumber\\
&+&\left.\left[(1-3\eta)(-14N_x^2
+6N_y^2)-(19-3\eta)\right](p_y^2-q_y^2)\right)\ ,\nonumber\\
h_+^{1SO}&=&-\frac{1}{r^2}\left[({\bf \Delta{\mbox{\boldmath$\cdot$}}q})p_x
+({\bf \Delta{\mbox{\boldmath$\cdot$}}p})q_x\right]\ ,\nonumber\\
h_+^{1,5}&=&\left(\frac{M}{r}\right)^{5/2}\left(\left[\frac{\delta m}{M}(1-2\eta)
\left(-\frac{37}{8}N_x^2+\frac{1}{4}N_y^2\right) -\frac{101-12\eta}{24}\right]N_y(p_x^2-q_x^2)\right.\nonumber\\
&+&\left[\frac{\delta m}{M}(1-2\eta)\left(-\frac{65}{12}N_x^2+\frac{15}{2}N_y^2\right)
-\frac{149-36\eta}{12}\right]N_x(p_xp_y-q_xq_y)\nonumber\\
&+&\left.\left[\frac{\delta m}{M}(1-2\eta)
\left(\frac{29}{4}N_x^2-N_y^2\right)+\frac{19-4\eta}{4}\right]N_y(p_y^2-q_y^2)\right)\ , \nonumber\\
h_+^{1,5SO}&=&\frac{2v}{r^2}\bigg\{3(2S_z+\Delta_z)(p_x^2-q_x^2)-\frac{1}{2}[(9{\bf S}
+5{\bf\Delta})\times(p_x{\bf p}-q_x{\bf q})]_y-[({\bf S}+{\bf\Delta})\times(p_y{\bf p}-q_y{\bf q})]_x\nonumber\\
&-&({\bf S}+{\bf\Delta}){\mbox{\boldmath$\cdot$}}\left[N_y(p_x{\bf q}+q_x{\bf p})
+N_x(p_y{\bf q}+q_y{\bf p})\right]\bigg\}\ ,
\end{eqnarray}
and
\begin{eqnarray}
h_{\times}^N&=&2\left(\frac{M}{r}\right)[-p_xq_x+p_yq_y]\ ,\nonumber\\
h_{\times}^{0,5}&=&\left(\frac{M}{r}\right)^{3/2}\frac{\delta m}{M}\left[N_yp_xq_x+3N_x(p_xq_y+q_xp_y)
-2N_yp_yq_y\right]\ ,\nonumber\\
h_{\times}^{1}&=&\left(\frac{M}{r}\right)^2\frac{1}{3}\left(\left[(1-3\eta)(10N_x^2-2N_y^2)
+ (19-3\eta)\right]p_xq_x-16(1-3\eta)N_xN_y(p_xq_y+q_xp_y)\right.\nonumber\\
&+&\left.\left[(1-3\eta)(-14N_x^2
+6N_y^2)-(19-3\eta)\right]p_yq_y\right)\ ,\nonumber\\
h_{\times}^{1SO}&=&-\frac{1}{r^2}\left[({\bf\Delta{\mbox{\boldmath$\cdot$}}q})q_x
-({\bf\Delta{\mbox{\boldmath$\cdot$}}p})p_x\right]\ ,\nonumber\\
h_{\times}^{1,5}&=&\left(\frac{M}{r}\right)^{5/2}\left(\left[\frac{\delta m}{M}(1-2\eta)
\left(-\frac{37}{4}N_x^2+\frac{1}{2}N_y^2\right)-\frac{101-12\eta}{12}\right]N_yp_xq_x\right.\nonumber\\
&+&\left[\frac{\delta m}{M}(1-2\eta)\left(-\frac{65}{12}N_x^2+\frac{15}{2}N_y^2\right)
-\frac{149-36\eta}{12}\right]N_x(p_xq_y+q_xp_y)\nonumber\\
&+&\left.\left[\frac{\delta m}{M}
(1-2\eta)\left(\frac{29}{2}N_x^2-2N_y^2\right)+\frac{19-4\eta}{2}\right]N_yp_yq_y\right)\ , \nonumber\\
h_{\times}^{1,5SO}&=&\frac{2v}{r^2}\bigg\{6(2S_z+\Delta_z)p_xq_x-\frac{1}{2}[(9{\bf S}+5{\bf\Delta})
\times(p_x{\bf q}+q_x{\bf p})]_y-[({\bf S}+{\bf\Delta})\times(p_y{\bf q}+q_y{\bf p})]_x\nonumber\\
&-&({\bf S}+{\bf\Delta}){\mbox{\boldmath$\cdot$}}\left[N_y(q_x{\bf q}-p_x{\bf p})
+N_x(q_y{\bf q}-p_y{\bf p})\right]\bigg\}\ .
\end{eqnarray}

The main steps of the method outlined above remain the same, only the expressions become simpler. The only relevant %%@
difference is that the equations of motion in this special case can be integrated in time and the explicit time %%@
dependence of the detectable waveform can be given. Although different post-Newtonian and spin-orbit contributions %%@
arise the orbital frequency, Eq. (\ref{circomega}), remains constant (up to 2.5\,PN order). As a consequence higher %%@
harmonics arise in the waveform compared to the Newtonian limit of this case, where the frequency of the emitted %%@
gravitational wave is twice the orbital frequency.

\section{Conclusions and outlook}

In this work we have discussed the orbital motion and the emitted gravitational radiation of compact binaries with %%@
the inclusion of the spin-orbit interaction and the relativistic post-Newtonian corrections up to 1.5\,PN order. We %%@
have given the steps of a parameterization independent method for the calculation of the polarization states of the %%@
detectable gravitational waveform. All the limiting cases, i.e. binaries with one spinning component, the extreme %%@
mass ratio limit and the circular orbit case can be calculated from this general description.

As a new result, the spin-orbit contributions to the equations of angular variables, including the spherical angles %%@
of spin vectors, are given. The usual 1\,PN corrections to the equations of motion are in agreement with previous %%@
calculations \cite{BFP}. To evaluate the lowest order contributions to the waveform it is usual to substitute the %%@
integrated equations of motion to the quadrupole formula \cite{ACST,Gop2}. Here we have considered the contributions %%@
from higher multipole moments up to 1.5\,PN order with the inclusion of spin-orbit terms.

In the post-Newtonian approximation physical systems can be studied in two steps. Since the detectable gravitational %%@
waveform can be unambiguously expressed as a linear combination of $h_+$ and $h_{\times}$ the first step is the %%@
construction of a method for the calculation of the polarization states which is independent from the %%@
parameterization of the orbit. To do this the full description of motion, including spin precession and the evolution %%@
of the angular variables, is required.

The second step determines our future plans in this area. We plan to discuss the explicit parameter dependence and %%@
the general structure of the waveform with the help of the generalized true anomaly parameterization \cite{KMG}. To %%@
analyze the effects of the eccentricity of the orbit we will investigate the circular orbit limit, where the results %%@
can be expressed explicitly in time. Furthermore, with the investigation of the one spinning and non-rotating limits %%@
we show the importance and the details of the spin-orbit interaction.

In the general description the next step is the evaluation of the spin-spin contributions at 2\,PN order, which %%@
provides corrections to both the dynamics of the binary and the detectable gravitational waves. This interaction is %%@
relevant for black hole-black hole binaries, but if one or two objects are neutron stars another effect, the %%@
quadrupole-monopole interaction, becomes important at 2\,PN order. The method presented above can be proven useful in %%@
the investigation of these higher order contributions.

\section*{Acknowledgments}
This work was supported by the Hungarian Scientific Research Fund (OTKA) Grants No. NI68228 and No. F049429. The %%@
authors thank \'A. Luk\'acs for his valuable comments on the manuscript.

\end{document}